\documentclass[12pt,a4paper]{article}
\usepackage{indentfirst}
\usepackage[T1]{fontenc}
\usepackage[utf8]{inputenc}
\usepackage{graphicx}
\usepackage{amsmath}
\usepackage{amssymb}
\usepackage{amsthm}
\usepackage{amsfonts}
\usepackage{indentfirst}
\usepackage{latexsym}
\usepackage{multicol}
\usepackage{geometry}
\usepackage{verbatim}
\usepackage{setspace}
\usepackage{float}
\usepackage{siunitx}
\usepackage{listings}
\usepackage{appendix}
\usepackage[dvipsnames]{xcolor}
\usepackage{caption}
\usepackage{subcaption}
\usepackage{tikz}
\usepackage{tikz-3dplot}
\usepackage{ulem}
\usepackage{physics}
\usetikzlibrary{matrix}
\usetikzlibrary{decorations.markings}
\usetikzlibrary{shapes.geometric}
\usetikzlibrary{3d, calc}
\usepackage{hyperref}
\usepackage{graphicx} 
\usepackage{stix}
\usepackage{mathtools}
\usepackage{nameref}
\usepackage{eqparbox}
\usepackage{braket}
\usepackage{mdframed}
\usepackage{ifthen}
\usepackage{alphalph} 

\newgeometry{tmargin=2cm, bmargin=2cm, lmargin=2cm, rmargin=2cm} 

\usepackage[parfill]{parskip} 

\theoremstyle{definition}

\numberwithin{equation}{subsection}

\let\oldsection\section
\renewcommand{\section}{%
  \renewcommand{\theequation}{\thesection.\arabic{equation}}
  \oldsection
}

\let\oldsubsection\subsection
\renewcommand{\subsection}{%
  \renewcommand{\theequation}{\thesubsection.\arabic{equation}}
  \oldsubsection
}

\renewcommand{\theequation}{\thesubsection.\arabic{equation}} 
\renewcommand{\theequation}{\thesubsection.\arabic{equation}\alphalph{\value{equation}}} 

\newcommand{\appendixnumbering}{%
  \renewcommand{\thesection}{\Alph{section}} 
  \renewcommand{\thesubsection}{\thesection.\arabic{subsection}} 
  \renewcommand{\theequation}{\thesubsection.\arabic{equation}} 
  \renewcommand{\theequation}{\thesubsection.\alphalph{\value{equation}}} 
}

\begin{document}
\pagenumbering{arabic}
\setcounter{page}{1}

\title{Ferromagnetic ordering in Hubbard models}
\author{Wojciech Niedzi\'o\l ka and Jacek Wojtkiewicz,\\
{\it Faculty of Physics, University of Warsaw,}\\
{\it Pasteura 5, 02-093 Warszawa, Poland}\\
(e-mails: ${\rm Wojciech.Niedziolka@fuw.edu.pl}$, ${\rm  wjacek@fuw.edu.pl}$)}
\date{}
\maketitle

\abstract{
One of the long-standing and only partially solved problems of theoretical condensed matter physics and mathematical physics is to demonstrate that ground states of some of the versions of the Hubbard model can exhibit a ferromagnetic ordering.
It has long been speculated that the opportunity crucial for the occurrence of ferromagnetism 
  is the structure of the lattice on which the Hubbard model is formulated \cite{TasakiMB}. As a consequence, while on simple cubic lattices no ferromagnetic ordering seems to be possible, it can naturally arise, even for low densities of magnetic moment carriers, on so-called frustrated lattices.
 
We investigate the problem of ground state ferromagnetic ordering with the use of
the formula for ground-state energy of interacting fermions as the first term of `density expansion', proven rigorously by Lieb, Seiringer and Solovej \cite{fermi exact} in continuum and by Giuliani \cite{hub exact} for the simple cubic lattice. Assuming that analogous expansion holds also for certain another lattices 
we apply this formula to five frustrated lattices --  among them to the face-centered cubic one.
The hypothesis is confirmed: most of examined models formulated on frustrated lattices do indeed have ferromagnetic ground states already for densities being moderate or even low.
Although the approach adopted cannot be treated as a rigorous proof that the ground state is ferromagnetic, the results obtained here strongly indicate that it can be the case. Moreover, as in some cases FM occurs at low densities, one can hope that it would be possible to prove convergence of the density expansion and prove rigorously the occurrence of `wealthy ferromagnetism' in these cases.
}






\section{Hubbard model -- Introduction: Historical context and rigorous results related to ferromagnetism}

The explanation of the {\em ferromagnetism} phenomenon has been the point of interest for a long time.
For a concise introduction to the whole subject, see \cite{Mattis}.
Magnets can be roughly divided into two classes: in first of them,   spins are localized and in the second --  particles carrying spin are itinerant. Models with  spins localized at lattice sites (most popular are Ising and Heisenberg models) give a good description of orderings only in insulators. But they give no satisfactory explanation for ferromagnetism in metals, where mobile electrons are responsible for this phenomenon.

The Hubbard model was proposed independently in 1963 by J. Hubbard \cite{Hubbard}, M. Gutzwiller \cite{Gutz} and J. Kanamori \cite{Kana}, as a model of interacting itinerant electrons in a solid. 
Initially the model was introduced as a way to explain ferromagnetism in metals. 
It is one of the simplest models that takes into account both the kinetic energy of electrons and their interaction, though in very simplistic form. 
Only electrons that occupy the same atom interact. 
Such an interaction may be considered a very simplified description of screened Coulomb interaction. 
The Hamiltonian is written down in Sec. \ref{sec:formulation}.

While the Hubbard model doesn't quantitatively describe any real material \footnote{although it can describe atomic gases in optical traps}, as during its derivation too many simplifications are made, it is a great tool for understanding many physical phenomena:
Metal-Insulator transition\cite{met-ins}, high-temperature superconductivity \cite{supercond}, Bose-Einstein condensation of cold atoms \cite{cold} and many more \cite{tasaki more} can be understood on the level of Hubbard model without the need for complicated Hamiltonians.
The Hubbard model can also be a model of magnetism \cite{tasaki}, which will also be the goal of this work.

There are few exact results for Hubbard model\footnote{The number of rigorous results for the Hubbard model is not large. 
The Hubbard model is difficult in dealing with -- `notoriously difficult', as it was said by  great E. Lieb\cite{Lieb}. It was more than thirty years ago, and the situation didn't change essentially in past years.}. 
Some important results, significant for the perspective of this work  are: Nagaoka's theorem \cite{nagaoka}, \cite{tasaki}; Lieb ferrimagnetism;\cite{Lieb ferri}; flat-band ferromagnetism \cite{Mielke}, \cite{tasaki}. These results are very important but somewhat unphysical (or very specific and therefore not general, like the kagome lattice) and 
so it is disputable in what extent they are significant for  what is usually called 'wealthy ferromagnetism' \cite{tasaki}:
Namely for physical range of densities and lattices.
To the best authors knowledge, rigorous attempts in this direction are very rare in literature \cite{TasakiAlmostFlat}, \cite{TanakaTasaki}. 

 On the other hand, many less-rigorous approaches have been applied to examine the itinerant ferromagnetism phenomenon. For interacting electron gas, important contributions are  early rough (mean-field type) considerations:\cite{Lenz}, \cite{Bloch},  \cite{Stoner}, followed by more precise \cite{LeeYang}. (Another approaches used resolvent method \cite{hub cubic} or DMFT \cite{FCC} methods). They can be viewed as  'density expansion', where one  finds the energy of the ground state as a function of densities of particles with spin up and spin down: $E(\rho_+, \rho_-)$.
The first two terms of this expansion were (almost half century late) rigorously proven both for the continuum case \cite{fermi exact}, \cite{FermiExact2} (recently even the third term has been found \cite{KannoRigorously}) and the Hubbard model \cite{hub exact}, , \cite{SeiringerYin}:
\[
E(\rho_+, \rho_-) = e_0(\rho_+) + e_0(\rho_-) + C a \rho_+ \rho_- + \cal{O}(\rho_+, \rho_-)
\]
where $e_0(\rho)$ is the energy of non-interacting fermions as a function of density, $C$ is model-dependent constant  and $a$ is the scattering length for two-particle scattering.
This formula is exact, but after neglecting the last (remainder) term, is an approximate one and applies only for low densities, so one cannot determine if the ground state is ferromagnetic with absolute certainty.
For example, in a continuum one can see ferromagnetism, but only in the {\em nonperturbative} regime -- for a density of the order of one.
In the Hubbard model in the perturbative regime, when $\rho\to0$ ferromagnetism {\em does not} appear \cite{hub exact} (this paper, however, does not exclude the possibility that it may appear for low, but finite, densities).

Let us also underline the fact that in the continuum, the zeroth term $e_0\sim\rho^{5\slash{}3}$ has standard form and it cannot be modified in any way.
The situation is vastly different in lattice models: here the structure of the lattice and therefore the dispersion relation is closely related to this term and will be different for different lattices.

On the other hand, on the heuristic level, an important factor that favors ferromagnetism is {\it high density of states near the Fermi level}. 
This idea is a content of the famous Stoner criterion\cite{Stoner}.
In the case of lattice models, density of states is strongly related to dispersion relation and geometry of the lattice. 
The density of states is (roughly speaking) an inverse of the derivative of energy with respect to momentum, so high densities of states should correspond to flatter dispersion relations. We wanted to check whether some analog of Stoner criterion holds also beyond mean-field level. 
So we state the main focus of the paper. 
It is a {\em study of ferromagnetism in the Hubbard model on frustrated lattices for low densities of fermions}.
To the best of our knowledge, this has 
not been studied from the point of view of density expansion beyond the mean-field level.

Our result is that for strongly frustrated systems, ferromagnetism {\it appears for low densities}, so one can expect that inclusion of higher order terms will not change its occurrence.

Our hypothesis is that the `tail' of the expansion can be estimated, and one could prove ferromagnetism in a rigorous setting.

Let us make also a remark: for half-filling and for models defined on bipartite lattices, one expects {\em antiferromagnetic} ordering (it also has not been proved, but were confirmed by many approximate methods, see for instance \cite{Hirsch}.)
This is not immediately related to our study, which is about ferromagnetism, but it is nevertheless an important observation that orderings in the Hubbard model differ drastically in two opposite cases: low densities (where we expect ferromagnetism) and half-filling (where we expect antiferromagnetism).
As one is expected that at half-filling we have antiferromagnetism, this gives a natural bound on the applicability of expansion in densities, which is the main tool used in this paper.

The organization of the paper is as follows. 
In Sec. \ref{sec:formulation}, the Hamiltonian of the model is written down and described. 
The core part of the paper is Sec.\ref{sec:Main}, where the ground state energy is computed as a function of densities for different lattices, and the consequences of those results for the nature of the ground state are studied.
The most important result is that {\em for frustrated lattices ferromagnetism appears at quite low densities}.
Sec.\ref{sec:Summa} summarizes results and discusses the perspective of further research.
An appendix is devoted to some technical aspects of calculations.

\section{Model formulation}
\label{sec:formulation}

We consider the Hubbard model in general form

\begin{align}
    H = -\sum_{(i, j), \sigma}^{} t_{i,j} (\hat{c}^\dagger_{i, \sigma} \hat{c}_{j, \sigma} + \hat{c}^\dagger_{j, \sigma} \hat{c}_{i, \sigma}) + U \sum_{i}^{} \hat{n}_{i, +} \hat{n}_{i, -}
\end{align}

where $\hat{c}_{i, \sigma}$, $\hat{c}^\dagger_{i, \sigma}$ are creation/annihilation operators of fermions with spin $\sigma= \pm$ on site $i$; those are fermionic operators satisfying usual anticommutation relations:
\[
 \left\{ \hat{c}^{\phantom{\dagger}}_{i, \sigma}, \hat{c}^\dagger_{j, \sigma'}\right\} = \delta_{i,j} \delta_{\sigma, \sigma'},
 \quad
 \left\{ \hat{c}^{\phantom{\dagger}}_{i, \sigma}, \hat{c}_{j, \sigma'}\right\} = 
 \left\{ \hat{c}^\dagger_{i, \sigma}, \hat{c}^\dagger_{j, \sigma'}\right\}
 = 0.
 \]
The Hamiltonian is composed of two parts: the kinetic term that describes hopping of fermions between lattice sites and a potential term that describes interaction of fermions on a site. 
The first summation goes over pairs of connected sites on the lattice, which depends on the geometry of the lattice with parameters $t_{i,j}$ called hopping matrix elements. 
The second summation simply goes over all lattice sites and is parametrized by $U$.

The model can be derived from the Schrödinger equation using many approximations; this was the approach of the original work \cite{Hubbard} of Hubbard. 
Such derivation gives us relation between $t, U$ parameters of the model and the wave functions of atoms.

\section{Ground state energy and magnetization of Hubbard models within density expansion}
\label{sec:Main}
\subsection{Ground state energy of interacting Fermi gas in continuum}

Before moving to the 3D Hubbard model, let us consider interacting Fermi gas and its ground state energy. 
Let us consider $N$ indistinguishable fermions of mass $m$ in a box of side length $L$ with periodic boundary conditions. 
The states are labeled by three quantum numbers, which define momentum $(k_x, k_y, k_z)$ and energy $e_{k_x, k_y, k_z} = \frac{\hbar^2}{2 m} (k_x^2 + k_y^2+k_z^2)$.
As we are interested in the ground state, we want to sum the $N$ lowest energies.
We  consider the case of a mixture of two types of indistinguishable particles (e.g. $N_\uparrow$ particles with spin 1/2 and $N_\downarrow$ particles with spin down). 

Obtaining the ground state energy
formula for interacting fermions rigorously is still an open problem. 
However, approximate methods were used to obtain leading asymptotics for the dilute gas of fermions\footnote{High-density limit can also be treated, however by different methods.}. 
It turns out that the following formula holds for the ground state energy in the thermodynamic limit. Namely, 
for the system of $N_\uparrow, N_\downarrow$ spin-up and spin-down particles of mass $m$ interacting with positive, radial potential of finite range, for $\lim_{N,V\to\infty} \frac{N_{ \uparrow,\downarrow}}{V}$ being finite ( $ N=N_\uparrow+ N_\downarrow$):

  \begin{align}
    e(\rho_\uparrow, \rho_\downarrow) = \frac{\hbar^2 }{2m } \frac{3}{5} \left(6 \pi^2\right)^\frac{2}{3} (\rho_\uparrow^\frac{5}{3} + \rho_\downarrow^\frac{5}{3}) + \frac{\hbar^2}{2 m} 8 \pi a \rho_\uparrow \rho_\downarrow + \text{higher order terms in}\;\rho_\uparrow, \rho_\downarrow
  \end{align}
  \label{DensityExp}
where $a$ is the two-body 
s-wave
 scattering length of the interaction potential.

The first-order correction (i.e. the two first terms of the expression above) was obtained by W. Lenz, F. Bloch and E. Stoner \cite{Lenz, Bloch, Stoner}, however with the wrong constant. 
An expression with the correct constant appeared first in a paper by Lee and Yang \cite{LeeYang}. 
Almost half a century ago, the formula \ref{DensityExp} has been rigorously proved \cite{fermi exact}, \cite{FermiExact2}. Next terms were initially calculated in the $\rho_\uparrow = \rho_\downarrow$ case \cite{fermi approx1, fermi approx2, fermi approx3}.  
Kanno extended this result and obtained second-order term in the general case of non-equal densities \cite{Kanno}. 
This result has been rederived using Effective Field Theory \cite{ChWorder2} and extended to third order in \cite{ChWorder3}. 
A remarkable result is also the rigorous derivation of the Kanno result in \cite{KannoRigorously}.

If we put constraint $\rho = \rho_\uparrow + \rho_\downarrow=\rm{const}$, then we can express the non-interacting part of energy $E_0$ as a function of single variable, say $\rho_\uparrow$: $E_0\equiv E_0(\rho_\uparrow)$.
Minimum of $E_0(\rho_\uparrow)$ is achieved for
$\rho_\uparrow=\frac{1}{2}\rho$, 
i.e. for equal densities, which means that the non-magnetic state is preferred. 
On the other hand, the first-order interaction is a concave function of $\rho_{\uparrow} (\rho-\rho_{\uparrow})$ preferring ferromagnetic ordering. 
For various values of density, these two contributions try to realize their preferred ordering: for low density the non-magnetic state will dominate, but when we increase density, from a certain critical density, the ferromagnetic ordering will dominate.
This critical density equal to $\rho_{\rm cr}=\pi/24a^3$ is unfortunately so large that is not legitimate to neglect further order terms, so prediction of ferromagnetism within two terms only is doubtful.

 
Comparing cases of continuous and lattice systems, an important opportunity is that in continuum we have "only one" form of energy of non-interacting fermions, and we don't see phase transition in the physical range of densities.
But in the case of the Hubbard model on a lattice, the situation differs in a significant way.
Every lattice has 
its specific dispersion relation and subsequently a unique form of energy of a non-interacting system. In particular, one can construct frustrated lattices and examine how frustration and high density of states influences tendency to ferromagnetism. It turns out that one can find lattices   exhibiting a phase transition to a ferromagnetic state at 
quite low density. This is presented in the following subsections.

\subsection{Ground state energy of Hubbard model on simple cubic lattice}
A lattice analog of the formula (\ref{DensityExp})
 for the Hubbard model, has been proven by Giuliani \cite{hub exact} for the upper bound, and by Seiringer and Yin \cite{SeiringerYin} for the lower bound.
They considered model with nearest neighbor hopping on cubic lattice.

The derived formula is similar to that obtained for a Fermi gas in the continuum:

\begin{subequations}\label{low density}
  \begin{align}
      E(\rho_\uparrow, \rho_\downarrow) =  E_0(\rho_\uparrow, \rho_\downarrow) + 8 \pi a \rho_\uparrow \rho_\downarrow + (\text{higher order terms in}\; \rho_\uparrow, \rho_\downarrow)
      \label{eGSYhb}
      \\
      \text{where  } \hspace*{2 cm} 8 \pi a = \frac{U}{U \gamma+1}  \hspace*{1 cm} \gamma = \frac{1}{2} \int_{\abs{k_i}\leq \pi } \frac{d^3k}{(2\pi)^3} \frac{1}{ e(\vectorarrow{k}) - e_{\rm min}}
  \end{align}
  \label{eGSYhb_pomocnicze}
\end{subequations}
where $E_0(\rho_\uparrow, \rho_\downarrow)$ is energy for the system of non-interacting particles ( i.e. $U = 0$) expressed as a function of density, $a$ is scattering length of the potential defined in terms of the dispersion relation of our lattice $e(\vectorarrow{k})$ and its 
minimum 
 $e_{\rm min}$. (For the simple cubic lattice, considered in \cite{hub exact} and \cite{SeiringerYin}, the dispersion is given by the formula (\ref{DispersionNNN}) with $t_2=0$.)

{\em Remark.} Authors of \cite{hub exact} and \cite{SeiringerYin} stated that  their analysis can be generalized to `cases with different hopping terms'. However, in our opinion, one should be careful with such statements. The reason is that low-density limit is based on two-body scattering theory. There is no problem when the dispersion $\epsilon(\mathbf{k})$ is a function, possessing isolated minimum at $\mathbf{k}=\mathbf{0}$. Fundamental problems appear when the minima are not isolated \cite{DerGer}. Such situations are not exceptional in solid state physics: for many lattices, one observes energy bands completely flat in one or even more directions (`sawtooth lattice', kagome lattice \cite{tasaki}). For this reason, we limit ourselves only to such dispersion which possess the square minimum at $\mathbf{k}=\mathbf{0}$. We also assume that for such dispersions, the ground-state energy possess the form (\ref{eGSYhb})

{\em Remark.}Using his formula, Giuliani showed that for the Hubbard model on the cubic lattice, the total spin of the system is zero for $\rho \rightarrow 0$.
However, his results don't specify what happens at low but non-zero densities.

In our paper, we assume validity of ground-state energy formula (\ref{eGSYhb})  for chosen frustrated lattices (under condition that dispersions $\epsilon(\mathbf{k})$ possess non-degenerate minima at $\mathbf{k}=\mathbf{0}$), and we skip the remainder term. (In many cases, we argue that this is reasonable assumption.) Under these assumptions, we examine ground states and show that for many of them energy is minimized for non-equal densities $\rho_\uparrow $ and  $\rho_\downarrow $ -- thus ground states are ferromagnetic. 
\subsection{Description of the problem studied and tools used: Green's functions, densities of states}

The goal of this work is to study this formula in the spirit of the previous section. 
In the case of interacting fermion gas in continuum, we had $e(\vectorarrow{k})\sim k^2$ dispersion relation. 
In the case of lattice models, we have much more freedom, as different geometries of lattice will result in different dispersion relations. 
We will focus on frustrated lattices to verify the hypothesis that frustration in Hubbard model aids ferromagnetism and to obtain quantitative results. 
In particular, frustration could lead to ferromagnetism at low density, unlike in free gas, where phase transition occurred at high density, which makes our low-density approximation inaccurate.

The energy without interaction splits into energies of spins up and spins down that are independent of each other $E_0(\rho_\uparrow, \rho_\downarrow) = E_0(\rho_\uparrow) + E_0(\rho_\downarrow)$.
At this point we encounter technical difficulty.
Energy $E(\rho)$ is given implicitly as $E(k_\rho)$.
For free fermion gas we have isotropy, and therefore one can easily write a formula for density as a function of Fermi momentum and vice versa - during calculations everything is expressed by $k=||\mathbf{k}||$.
As a consequence, writing the energy of non-interacting particles as a function of their density, i.e. the formula for $E(\rho)$, is a straightforward and easy task.
On the contrary, when we consider the Hubbard model, the situation gets complicated: both quantities, $E=E(\mathbf{k})$ and $\rho=\rho(\mathbf{k})$, are {\em not} isotropic.
In such a situation, expression of $E$ as a function of $\rho$ is not straightforward.
We solve this problem by the use of {\em density of states}. 
It can be computed using {\em Green's functions} \cite{economu}, which give straightforward formulas that are easy to work with in numerical calculations.

Recall that the Green's function $G(z)$, $z\in \mathbb{C}$ is given by

\begin{align}
    G(z) = \frac{1}{(2 \pi)^d} \int_{1\text{BZ}} \frac{d^d k}{z - e(\vectorarrow{k})}
\end{align}

Then the density of states $D(E)$ can be found from the forward Green's function $G^+(z)$ \cite{economu} 

\begin{subequations} \label{def density}
  \begin{align}
      G^+(z) = \lim_{\epsilon \rightarrow 0} G(z+i \epsilon)\\
      D(z) = - \frac{1}{\pi} \mathrm{Im} G^+(z)
  \end{align}
\end{subequations}

These formulas allows for straightforward numerical calculation of density of states once we have dispersion relation, so first we have to find them for lattices of interest, and then we can use the formulas and obtain desired results.

Density of states is calculable for the simple cubic lattice as an integral over elliptic integrals \cite{economu}, \cite{hub cubic}.
However, calculations are no longer possible in an explicit manner in the case of more complicated dispersion relations than that of the simple cubic lattice. 
As we are interested in frustrated lattices, those calculations need to be performed fully numerically. 
It is good to first see how the numerical approach works for a lattice that has known solution.

\begin{figure}[H]
    \centering
    \includegraphics[width = 12 cm]{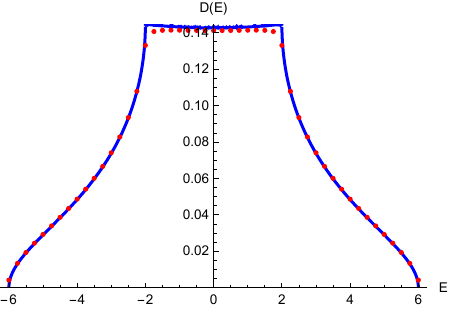}
    \caption{Density of states for 3D, numerical results (red points) compared to exact result (blue line)}\label{3D exact vs numerics}
\end{figure}

The definition of density of states has a limit with $\epsilon$ going to 0 in expression (\ref{def density}), when computing numerically density at some $E$ we can find the value of Green's function at $E+i \epsilon$ for only non-zero $\epsilon$, and we know that it will continuously approach $G^+$ as $\epsilon$ goes to zero. 
For this reason calculations are performed with a finite, but small value of $\epsilon$ for many energies from the spectrum to obtain a list of points ($E$, $D(E)$), which in further computations will be interpolated to obtain the density of states.

After calculations of DOS, we use it to calculate the dependence of the ground state energy for a given value of fraction $\frac{\rho_\uparrow}{\rho}$. 
The procedure is as follows. 
We use density of states $D(e)$ to find energy of non-interacting particles $E_0$ and density of particles $\rho$ for a single component.
In our case, all states from $e_{min}$ up to certain energy $e_F$ are occupied, then 

\begin{subequations}
    \begin{align}
        E_0(e_F) &= \int^{e_F}_{e_{min}}  e D(e)de\\
        \rho(e_F) &= \int^{e_F}_{e_{min}}  D(e)de
    \end{align}
\end{subequations}

In the next step, we invert the relation $\rho(e_F)$ to obtain $e_F(\rho)$ and express $E_0$ in terms of $\rho$.
Numerically this is done by computing pairs $(\rho(e_F), E_0(e_F))$ for many $e_F$, which gives us points function $E_0(\rho)$.
We calculate those pairs for many $e_F$ and then interpolate the results.

Finally, the energy of a mixture of two types of particles (with spin up and spin down) is $E_0(\rho_\uparrow, \rho_\downarrow) = E_0(\rho_\uparrow) + E_0(\rho_\downarrow)$. 
After calculating $\gamma$ for this given lattice (analogously as in the formula ( \ref{low density})) and corresponding scattering length $a$,
we are ready to analyze the problem.

\subsubsection{Critical density of states}

In our approximation the total energy is given by formula \ref{eGSYhb}, we analyze various lattices by $E$ as a function of $\rho_\uparrow/\rho$ and observe how this plot changes when we change $\rho$. 
We look for phase transition i.e. $\rho_c$ for which the minimum of energy goes from zero magnetization to non-zero magnetization; in our picture this means that $\rho_\uparrow/\rho=1/2$ is no longer global minimum of energy $E$.
It turns out that we can derive the condition for $\rho_\uparrow/\rho=1/2$ to no longer be local minimum (for details of calculation see Appendix \ref{sec:exact second derivative}), we introduce the notion of critical density of states $D_{crit}$ and obtain condition when $\rho_\uparrow/\rho=1/2$ is local minimum

\begin{align}
    D\left(e_F\left(\frac{1}{2}\rho\right)\right) < D_{crit} =  \frac{1}{8 \pi a}
\end{align}

where $D(e)$ is density of states and $e_F(\frac{1}{2}\rho)$ is Fermi energy corresponding to half of our total density $\rho$.
This condition gives us implication only in one direction, if our density of states is bigger that critical density of states, then we have local maximum and conclude that ground state for this value of $\rho$ has non-zero magnetization.
If our density of states is lower that critical value we know that $\rho_\uparrow/\rho=1/2$ is a local minimum, but know nothing about global minimum and nature of ground state, which in the case of second order phase transition can be magnetized.
Nevertheless this formula gives us insight into when we can expect phase transition to happen, for a given profile of density of states and chosen $u$ we can plot line at $D_{crit}$. It may intersect density of states and certain energies $e_F$; we then know that at $\rho_c=2 \rho(e_F)$ we will have change of sign of second derivative at $\rho_\uparrow / \rho=1/2$.
The results presented further agree with this, critical density of states gives accurate prediction when phase transition is of the first order, but for second order phase transition gives us little information, as $\rho_\uparrow / \rho=1/2$ is still local minimum when global minimum with non-zero magnetization forms.

\subsubsection{Simple cubic lattice}

We begin with reproduction of known results for the Hubbard model on a simple cubic lattice.
We have found that the ground state possesses zero magnetization in the whole range of physical densities up to $\rho=0.8$ and Coulomb constant up to $U=20$.
It is consistent with existing results \cite{Obermeier_DMFT} (where FM within the DMFT method was not present up to $U=20$) as well as \cite{Mattis}, ch. 4.15, \cite{KotliarRuckenstein} (where FM was absent up to $ U \approx 15$).
Above $U=70$, we do observe phase transition and ferromagnetic ordering (at $U=70$, it happens around $\rho=0.5$). However such high values of the $U$ parameter are unphysical and not relevant to our considerations, where we want to investigate physically relevant systems. 

Density of states is presented on Fig. \ref{3D exact vs numerics}; it has maximal value around $0.141$, meanwhile critical density of states for $U=20$ has value $0.17$. The value
$U=70$ corresponds to $a=0.283$ and critical density of states, now equal to $0.14$, can be achieved which implies that phase transition must happen.

\begin{figure}[H]
    \centering
    \begin{subfigure}{0.35\textwidth}
        \centering
        \includegraphics[width=\linewidth]{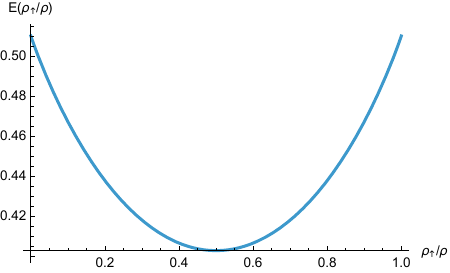}
    \end{subfigure}
    \hspace*{1.5cm}
    \begin{subfigure}{0.35\textwidth}
        \centering
        \includegraphics[width=\linewidth]{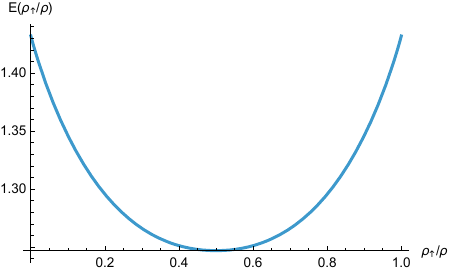}
    \end{subfigure}
    \\
    \begin{subfigure}{0.35\textwidth}
        \centering
        \includegraphics[width=\linewidth]{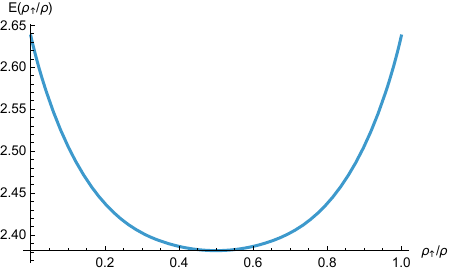}
    \end{subfigure}
    \hspace*{1.5cm}
    \begin{subfigure}{0.35\textwidth}
        \centering
        \includegraphics[width=\linewidth]{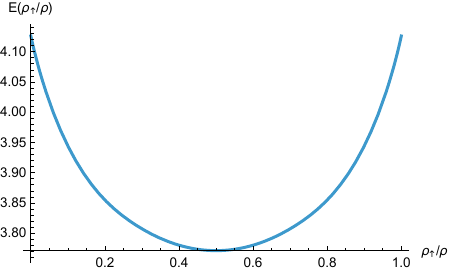}
    \end{subfigure}
    
    \caption{Plot of $E(\rho_\uparrow/\rho)$ for $U=20$ corresponding to $a=0.23$ and $\rho=0.2, 0.4, 0.6, 0.8$. As we increase density for cubic lattice we see that the ground state has no spontaneous magnetization up to non-physical range $\rho=0.8$}
\end{figure}

\subsubsection{Next nearest neighbors (NNN) hoppings}\label{NNN}

The second lattice (more precisely, the one-parameter family of lattices) that was considered was a cubic lattice with additional hoppings $t_2$ between site $(i_x, i_y, i_z)$ and sites $(i_x\pm 2, i_y, i_z)$,  $(i_x, i_y\pm 2, i_z)$,  $(i_x, i_y, i_z\pm 2)$. 

\begin{figure}[H]
    \centering
    \includegraphics[width=5cm]{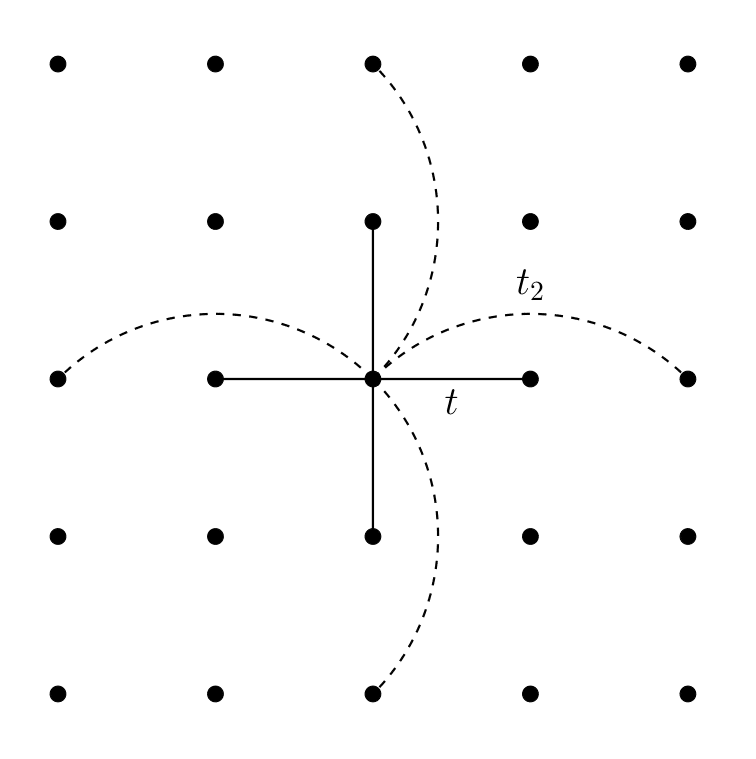}
    \caption{Next nearest neighbors hoppings visualized in the 2D case.
    Thick lines represent nearest neighbors hopping $t$ and dashed lines represent next nearest neighbors hoppings $t_2$.}
\end{figure}

We find easily the dispersion relation (by the aid of discrete Fourier transform) as
\begin{align}
e_{\text{NNN}}(\vectorarrow{k})
=&- 2t \left( \cos{k_x} + \cos{k_y} + \cos{k_z} \right)  -  2 t_2 \left( \cos{2 k_x} + \cos{2 k_y} + \cos{2 k_z} \right)
\label{DispersionNNN}
\end{align}
After plugging it into the Green's function, we can obtain the density of states for given values of parameters $t, t_2$.
The scattering length $a$ is computed according to \ref{low density}.

We are interested in values of $t_2$ belonging to the interval from 0 to $t/4$. Value
$t_2=0$ corresponds to a simple cubic lattice, whereas the case $t_2=t/4$ corresponds to maximal frustration. 
In this case near the bottom, energy levels are topological spheres, but the low-momenta asymptotics of dispersion $e_{\text{NNN}}(\vectorarrow{k})$ is proportional to $k^4$ instead of the standard quadratic one.

\begin{figure}[H]
    \centering
    \begin{subfigure}{0.35\textwidth}
        \centering
        \includegraphics[width=\linewidth]{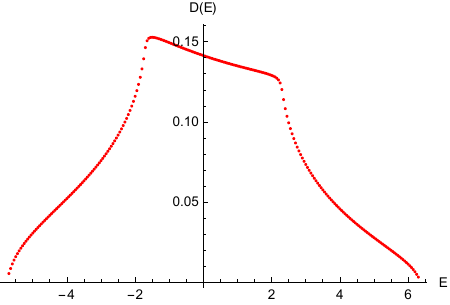}
    \end{subfigure}
    \hspace*{1.5cm}
    \begin{subfigure}{0.35\textwidth}
        \centering
        \includegraphics[width=\linewidth]{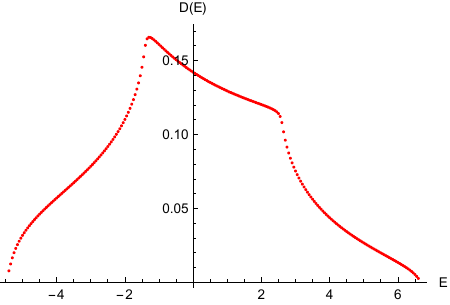}
    \end{subfigure}
    \\
    \begin{subfigure}{0.35\textwidth}
        \centering
        \includegraphics[width=\linewidth]{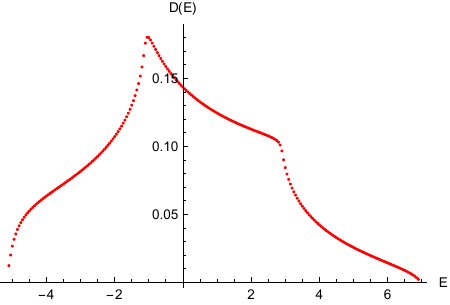}
    \end{subfigure}
    \hspace*{1.5cm}
    \begin{subfigure}{0.35\textwidth}
        \centering
        \includegraphics[width=\linewidth]{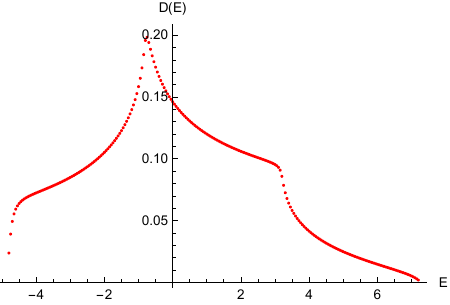}
    \end{subfigure}
    
    \caption{Density of states for $t=-1$ and $t_2 = -0.05, -0.10, -0.15, -0.20$.
    We can see the continuous transition from the shape for a simple cubic lattice to the shape for maximal frustration}
    \label{NNN gradual frustration}
\end{figure}

\begin{figure}[H]
  \centering
  \includegraphics[width = 12 cm]{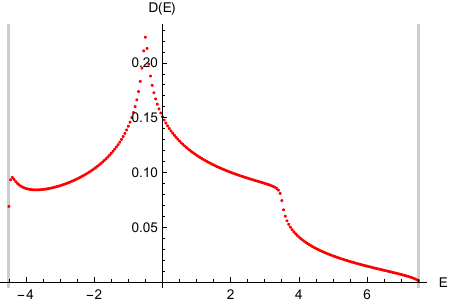}
  \caption{Density of states at maximal frustration $t=-1$ and $t_2=-0.25$}\label{NNN full frustration}
\end{figure}

The case of maximal frustration needs to be treated with caution; the low-momentum energy asymptotic is quartic, therefore one would have to investigate the zero-energy Schrödinger equation, from which we obtain scattering length.
However, it turns out that for values of $t_2$  that are close to 0.25, we obtain results that quantitatively are similar to the case of maximal frustration,

Below we present densities of state for $t=-1$ and $t_2 = -0.05, -0.10, -0.15, -0.20$ in figure \ref{NNN gradual frustration} and the case of maximal frustration $t_2=-0.25$ in Figure \ref{NNN full frustration}

In the case of non-maximal frustration, we see that near the boundaries of an interval the numerically calculated density tends to 0.

Considering the case of maximal frustration,
the left and right vertical lines are the lower and upper bounds of the spectrum at $e_{min}= -4.5$ and $e_{max}=7.5$ respectively, while the middle one is located at $e=0$. 
At $e= -4.5$, density tends to a finite value.
Around $e=-1/2$, we see singular behavior.
Note that the singularity is only present in the case of maximal frustration.
It is also quite far from the lower bound of the spectrum, compared to the two next lattices; because of that, here we can use $t_2=-0.25$ (the case of maximal frustration) without any problems. 

Once we have it, we can use it to the find dependence of energy on the number of fermions $E_0(\rho)$ and analyze how the ground state changes when we change $U$ (and consequently $a$) and $\rho$. 
We see that strong frustration leads to the emergence of spontaneous magnetization (unlike the simple cubic lattice; see the previous subsection): phase transition happens around $\rho = 0.6$. 
However, this phase transition can be questioned, as our approximation is valid only for low densities.

For $U=25$ and $a=0.31$ critical density of states is equal to $D_{crit}=0.128$ and can be achieved due to singular behavior in density of states.
However as singularity is present in middle of spectrum we expect that density corresponding to phase transition will not be small, which turns out to be the case.

\begin{figure}[H]
  \centering
  \begin{subfigure}{0.35\textwidth}
      \centering
      \includegraphics[width=\linewidth]{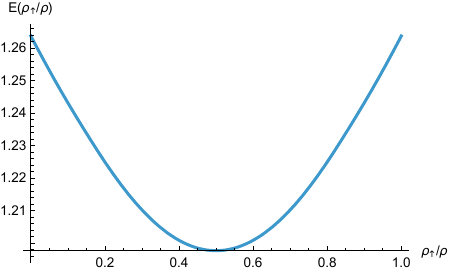}
      \caption{Initially for low $a$ and $\rho$ the ground state has magnetization 0 ($\rho=0.5$)}
  \end{subfigure}
  \hspace*{1.5cm}
  \begin{subfigure}{0.35\textwidth}
      \centering
      \includegraphics[width=\linewidth]{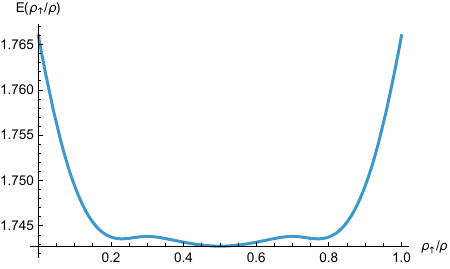}
      \caption{When we increase $\rho$, we observe that two new local minima form ($\rho=0.606$)}
  \end{subfigure}
  \\
  \begin{subfigure}{0.35\textwidth}
      \centering
      \includegraphics[width=\linewidth]{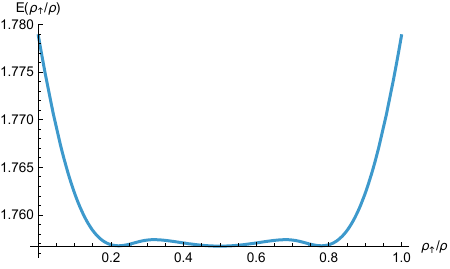}
      \caption{At some $\rho$ all 3 minima have the same value ($\rho=0.6085$)}
  \end{subfigure}
  \hspace*{1.5cm}
  \begin{subfigure}{0.35\textwidth}
      \centering
      \includegraphics[width=\linewidth]{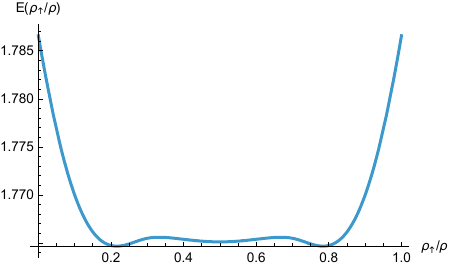}
      \caption{Increasing the $\rho$ any further results in phase transition. Now the ground state has non-zero  magnetization ($\rho=0.61$)}
  \end{subfigure}
  \\
  \begin{subfigure}{0.35\textwidth}
      \centering
      \includegraphics[width=\linewidth]{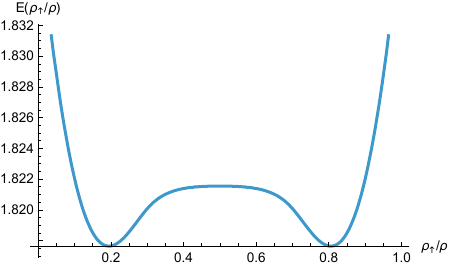}
      \caption{Increasing $\rho$ removes third minimum ($\rho=0.62$)}
  \end{subfigure}
  \hspace*{1.5cm}
  \begin{subfigure}{0.35\textwidth}
      \centering
      \includegraphics[width=\linewidth]{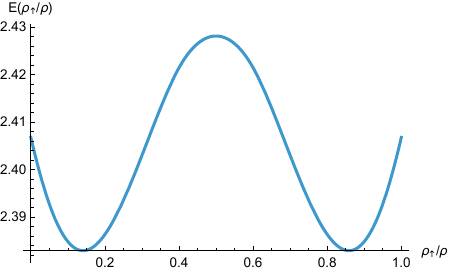}
      \caption{Finally the two minima move to two points with maximal achievable magnetization, further increase of $\rho$ decreases magnetization of ground state($\rho=0.72$)}
  \end{subfigure}
  \caption{First order phase transition in cubic lattice with NNN at $a = 0.31$ corresponding to $U = 25$, $t=-1$ and $t_2=-0.25$. The phase transition takes place at $\rho_c \approx 0.61$.}
\end{figure}



\subsubsection{Diagonal nearest neighbour (DNN) hoppings}\label{DNN}

The second lattice that was considered was a cubic lattice with additional hoppings $t_3$ between site $(i_x, i_y, i_z)$ and sites $(i_x\pm 1, i_y\pm 1, i_z)$, $(i_x\pm 1, i_y, i_z\pm 1)$, $(i_x, i_y\pm 1, i_z\pm 1)$.


\begin{figure}[H]
    \centering
    \includegraphics[width=5cm]{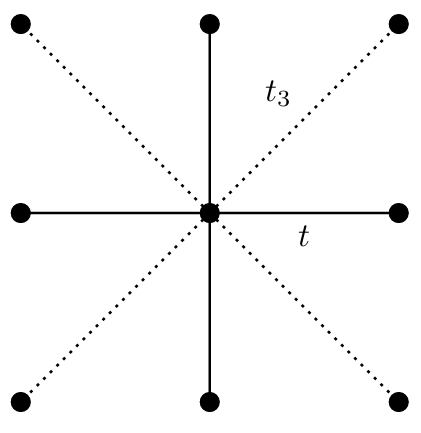}
    \caption{Diagonal nearest neighbors hoppings visualized in the 2D case.
    Thick lines represent nearest neighbors interaction $t$ and dashed lines represent diagonal nearest neighbors interactions $t_3$.}
\end{figure}

Again, we obtain the following dispersion relation by the aid of discrete Fourier transform:

\begin{align}
    e_{DNN}(\vectorarrow{k}) = - 2t \left( \cos{k_x} + \cos{k_y} + \cos{k_z}\right) -  t_3 \left( \cos{(k_x + k_y)} + \cos{(k_y + k_z)} + \cos{(k_z + k_x)} +\right. \\ 
    \left.\cos{(k_x - k_y)} + \cos{(k_y - k_z)} + \cos{(k_z - k_x)}\right) \notag
\end{align}

After plugging it into the Green's function, we can obtain the density of states for some specific parameters $t, t_3$.

\begin{figure}[H]
    \centering
    \begin{subfigure}{0.35\textwidth}
        \centering
        \includegraphics[width=\linewidth]{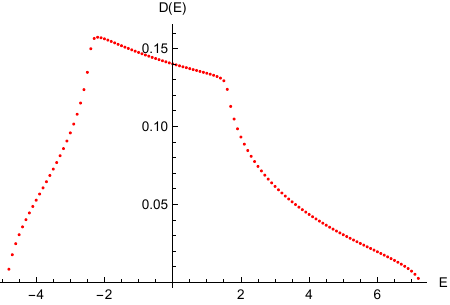}
    \end{subfigure}
    \hspace*{1.5cm}
    \begin{subfigure}{0.35\textwidth}
        \centering
        \includegraphics[width=\linewidth]{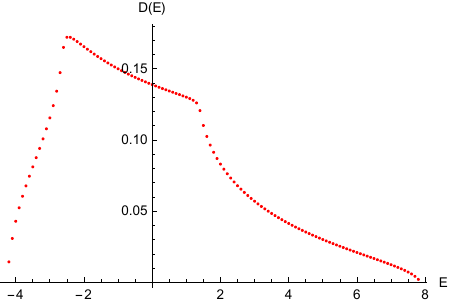}
    \end{subfigure}
    \\
    \begin{subfigure}{0.35\textwidth}
        \centering
        \includegraphics[width=\linewidth]{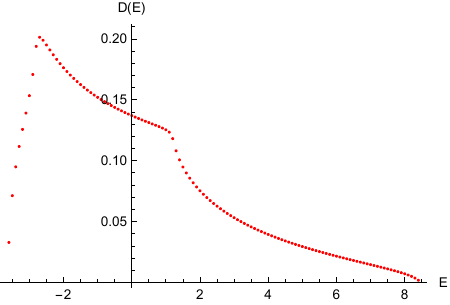}
    \end{subfigure}
    \hspace*{1.5cm}
    \begin{subfigure}{0.35\textwidth}
        \centering
        \includegraphics[width=\linewidth]{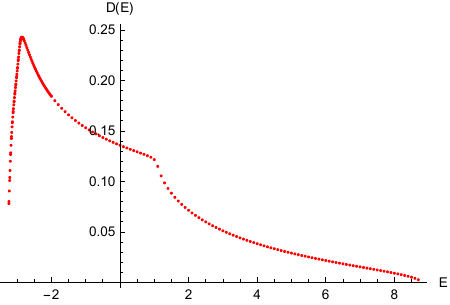}
    \end{subfigure}
        \\
      \begin{subfigure}{0.35\textwidth}
   	\centering
   	\includegraphics[width=\linewidth]{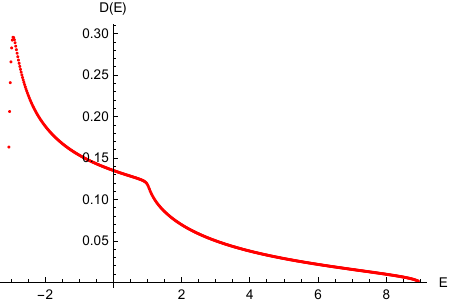}
   \end{subfigure}
    \caption{Density of states for $\frac{t_3}{t} = 0.10, 0.15, 0.20, 0.23, 0.245$. We can see the continuous transition from the shape for simple cubic lattice to the shape for maximal frustration}
\end{figure}

The case of maximal frustration $t_3=t/4$ is much more problematic than in the previous subsection.
The dispersion relation $E(\mathbb{k})$ in the previous subsection had minima at isolated points. Here, minima are not isolated -- they are located on whole lines $k_i=0$, $i=x,y,z$.
Here, in-depth analysis of the zero-energy Schrödinger equation is even more crucial, as we cannot a priori assume the same physical picture of scattering as in the case of non-degenerate maximum. However, we will make most computations for the almost maximally frustrated case, where those problems are not present. We present also the calculations  for {\em maximally} frustrated case, without justification of its validity -- for the sake of comparison.

\begin{figure}[H]
  \centering
  \includegraphics[width = 12 cm]{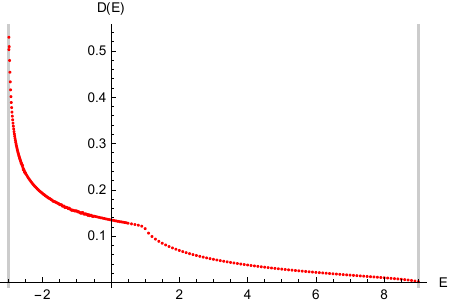}
  \caption{Density of states for the case of maximal frustration $t_3=t/4$. We observe singularity at the bottom.}
\end{figure}

It turns out that maximal frustration case differs in a significant way from the situation of near-maximal frustration.
The density of states for maximal frustration has a singularity  at the lower end of the spectrum at $E=-3$. 
An analysis of  the phase transition shows, surprisingly, that the ground state has {\em maximal magnetization for density tending to zero}. 
We observe similar behavior in flat-band ferromagnetism \cite{Mielke} \cite{tasaki}, where the ground state possesses saturated magnetization for certain range of densities.
In our case, for some larger density, the system exhibits a phase transition to zero magnetization. All together seems to be non-physical behavior. The root of this behavior is probably related to completely different form of the solution of scattering equation, and it is doubtful whether the formula (\ref{eGSYhb}) for low-density ground state energy is correct.

\begin{figure}[H]
	\centering
	\begin{subfigure}{0.35\textwidth}
		\centering
		\includegraphics[width=\linewidth]{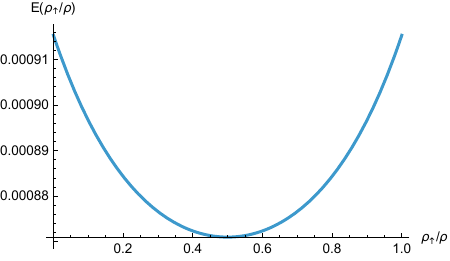}
		\caption{Initially for low $a$ and $\rho$ the ground state has magnetization 0 ($\rho=0.02$)}
	\end{subfigure}
	\hspace*{1.5cm}
	\begin{subfigure}{0.35\textwidth}
		\centering
		\includegraphics[width=\linewidth]{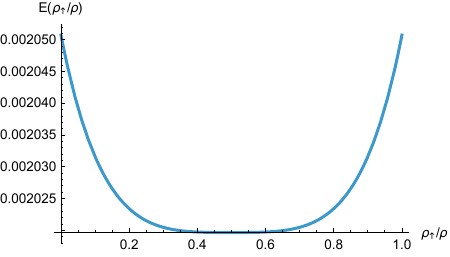}
		\caption{When we increase $\rho$, we observe flattening of total energy as function of magnetization ($\rho=0.031$)}
	\end{subfigure}
	\\
	\begin{subfigure}{0.35\textwidth}
		\centering
		\includegraphics[width=\linewidth]{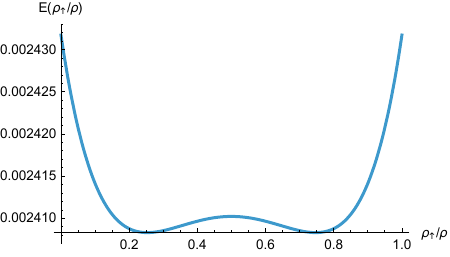}
		\caption{At some $\rho$ minimum splits into two minima with non-zero magnetization ($\rho=0.034$)}
	\end{subfigure}
	\hspace*{1.5cm}
	\begin{subfigure}{0.35\textwidth}
		\centering
		\includegraphics[width=\linewidth]{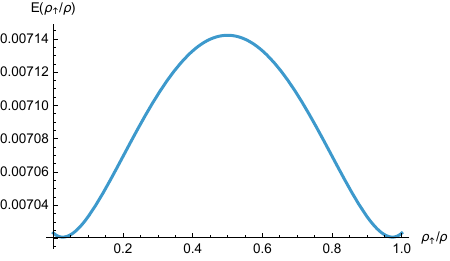}
		\caption{Increasing the $\rho$ increases magnetization of the ground state ($\rho=0.06$)}
	\end{subfigure}

	\caption{Second-order phase transition in an almost maximally frustrated ($t_3=0.245 t$) cubic lattice with DNN for $a = 0.14$ corresponding to $U = 5$. The critical density is quite low: $\rho_c=0.03$}
\end{figure}

If we consider "almost maximally frustrated lattice", after repetition of the calculations for $t_3=0.245 t$, we can resolve the singularity better and obtain physically expected results that show a proper picture of phase transition. It takes place at quite a low density $\rho=0.03$, where we observe an appearance of the  spontaneous magnetization. It is a phase transition of second order.

Intuition given by critical density of states explains behavior in this case.
For maximal frustration density of states is monotonically decreasing, so at low $\rho$ we are above $D_{crit}$ and non-magnetized state cannot be a ground state, for higher $\rho$ we are below $D_{crit}$ and non-magnetized state is local minimum of energy and only in this case can be ground state.
This explains that density of states corresponding to maximal frustration cannot give us physical phase transition and happens in opposite direction.
For $t_3/t<0.25$ density of states rapidly drops to zero when we approach bottom of spectrum, so for very low $\rho$ the ground state is non-magnetized.


\begin{figure}[H]
	\centering
	\begin{subfigure}{0.35\textwidth}
		\centering
		\includegraphics[width=\linewidth]{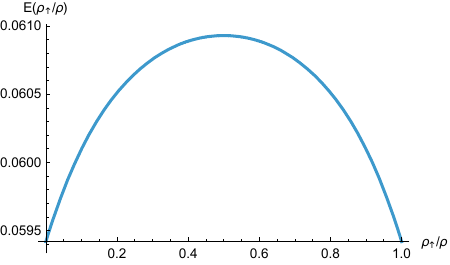}
		\caption{Initially for low $a$ and $\rho$ the ground state has  magnetization 1 ($\rho=0.2$)}
	\end{subfigure}
	\hspace*{1.5cm}
	\begin{subfigure}{0.35\textwidth}
		\centering
		\includegraphics[width=\linewidth]{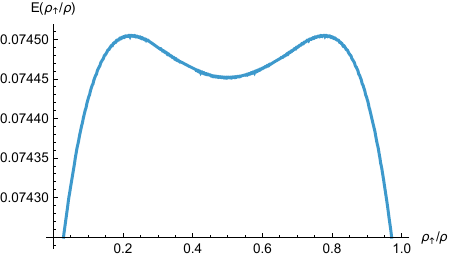}
		\caption{When we increase $\rho$, we observe new local minima that is formed at magnetization equal to 0 ($\rho=0.22$)}
	\end{subfigure}
	\\
	\begin{subfigure}{0.35\textwidth}
		\centering
		\includegraphics[width=\linewidth]{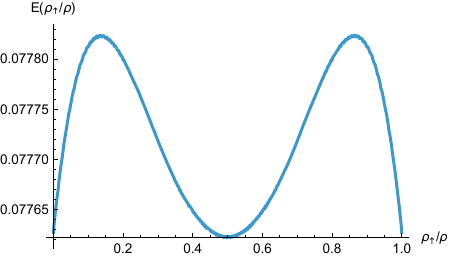}
		\caption{At some $\rho$ all 3 minima have the same value ($\rho=0.2244$)}
	\end{subfigure}
	\hspace*{1.5cm}
	\begin{subfigure}{0.35\textwidth}
		\centering
		\includegraphics[width=\linewidth]{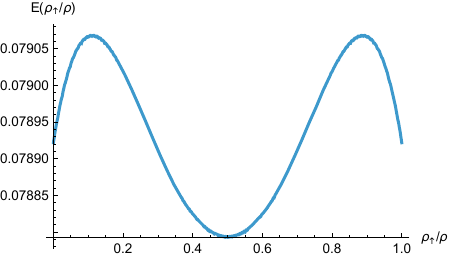}
		\caption{Increasing the $\rho$ any further results in phase transition, now the ground state has zero magnetization ($\rho=0.226$)}
	\end{subfigure}
	\\
	\begin{subfigure}{0.35\textwidth}
		\centering
		\includegraphics[width=\linewidth]{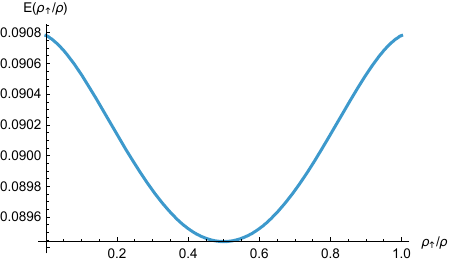}
		\caption{Increasing $\rho$ removes two minima with non-zero magnetization ($\rho=0.24$)}
	\end{subfigure}

	\caption{Non-physical behavior for completely frustrated  ($t_3=0.25 t$) cubic lattice with DNN for $a = 0.15$ corresponding to $U = 5$}
\end{figure}

\subsubsection{FCC lattice}

The most physically relevant lattice that was investigated was the FCC lattice \cite{FCC}, the face-centered cubic lattice, which can be defined as a cubic lattice with an additional site at every cube face. 
Alternatively, we can define it as a set of points $\{(i_x, i_y, i_z)|i_x+i_y+i_z\equiv 0 \mod{2}\}$. 

\begin{figure}[htbp]
    \centering
    \includegraphics[width=4cm]{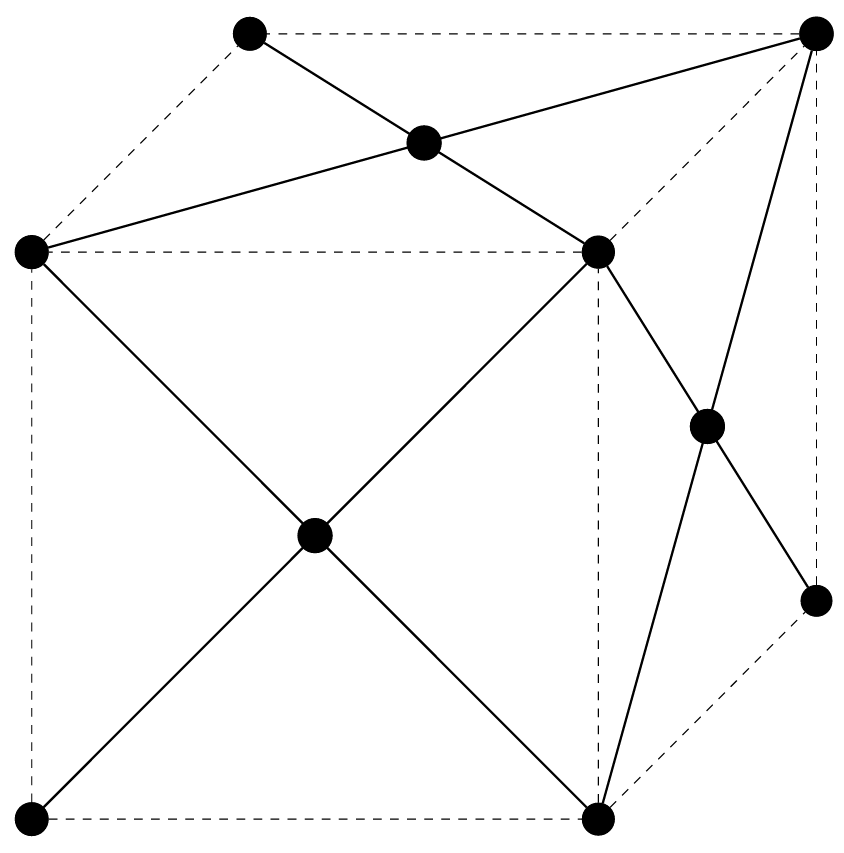}
    \caption{Single cell of FCC lattice. Solid lines correspond to hopping constant $t$ and dashed lines to $t_2$}
\end{figure}
To find the dispersion relation we write the Hamiltonian as

\begin{align}
    H^{\vec{i}}_{\vec{j}} = 
    \begin{cases}
    -t & \text{if }\abs{\vec{i}-\vec{j}}=\sqrt{2}\\
    -t_2 & \text{if }\abs{\vec{i}-\vec{j}}=2\\
    0 & \text{otherwise}
    \end{cases}
\end{align}
where $\vec{i}$ and $\vec{j}$ are vectors of integers that specify the point on the lattice. 
While it may seem initially that the lattice has two types of sites, the corners of cubes and points on the faces of cubes, all points have the same number of neighbors (12 nearest and 6 next nearest) that are positioned in exactly the same way.
As a result, we have a single equation to solve, and we obtain a one-band dispersion relation.
Analogously as in previous cases the dispersion is obtained by discrete Fourier transform
and after calculations similar to the last two cases we obtain 
\begin{align}
    e_{FCC}(\vectorarrow{k}) = - 4t \left( \cos{k_x} \cos{k_y} + \cos{k_y} \cos{k_z} + \cos{k_z} \cos{k_x}\right) -  t_2 \left( \cos{2 k_x} + \cos{2 k_y} + \cos{2 k_z}\right) 
\end{align}

After plugging it into the Green's function, we can obtain the density of states for some specific parameters $t, t_2$.

\begin{figure}[H]
    \centering
    \begin{subfigure}{0.35\textwidth}
        \centering
        \includegraphics[width=\linewidth]{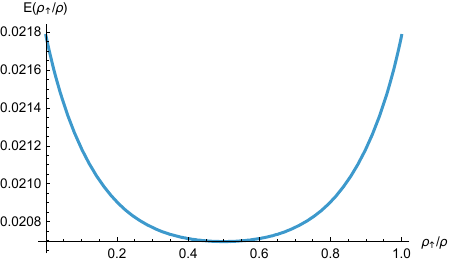}
        \caption{Initially for low $a$ and $\rho$ the ground state has magnetization 0 ($\rho=0.08$)}
    \end{subfigure}
    \hspace*{1.5cm}
    \begin{subfigure}{0.35\textwidth}
        \centering
        \includegraphics[width=\linewidth]{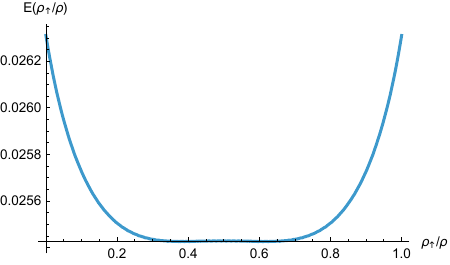}
        \caption{When we increase $\rho$, we observe that minimum flattens out ($\rho=0.09$)}
    \end{subfigure}
    \\
    \begin{subfigure}{0.35\textwidth}
        \centering
        \includegraphics[width=\linewidth]{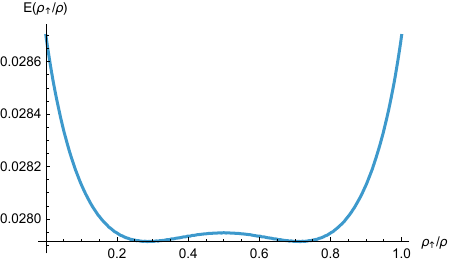}
        \caption{At some $\rho$ minimum splits into two with non-zero magnetization ($\rho=0.095$)}
    \end{subfigure}
    \hspace*{1.5cm}
    \begin{subfigure}{0.35\textwidth}
        \centering
        \includegraphics[width=\linewidth]{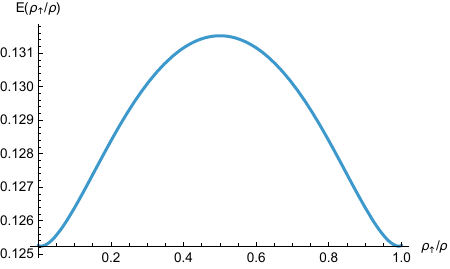}
        \caption{Increasing the $\rho$ increases magnetization of the ground state until at a certain point system reaches maximum magnetization ($\rho=0.23$)}
    \end{subfigure}
    \caption{Second-order phase transition in FCC lattice with $t=-1$ and $t_2 = -0.25$ for $a = 0.16$ corresponding to $U = 5$. The critical density is quite low: $\rho_c = 0.09$}
    \label{fig:fcc025}
\end{figure}

The situation is very similar to the previous case of DNN lattice.
For $t_2=t/2$ we have non-isolated critical points in dispersion - we have 'maximal frustration'.
Density of states has singularity of the form $\rho\sim 1/\sqrt{|E-E_0|}$.
The comments from the previous section are also relevant here. 
Below, we present  results obtained for $t_2=t/4$, i.e. not a  maximal frustration (Fig. \ref{fig:fcc025}).
The second phase transition happens at quite low density $\rho=0.088$.



\begin{figure}[H]
    \centering
    \begin{subfigure}{0.35\textwidth}
        \centering
        \includegraphics[width=\linewidth]{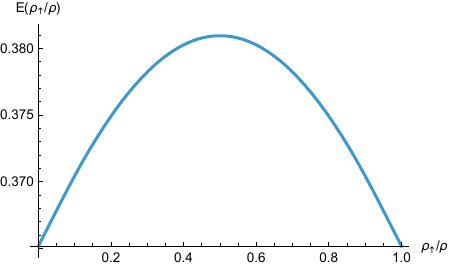}
        \caption{Initially for low $a$ and $\rho$ the ground state has maximum magnetization ($\rho=0.5$)}
    \end{subfigure}
    \hspace*{1.5cm}
    \begin{subfigure}{0.35\textwidth}
        \centering
        \includegraphics[width=\linewidth]{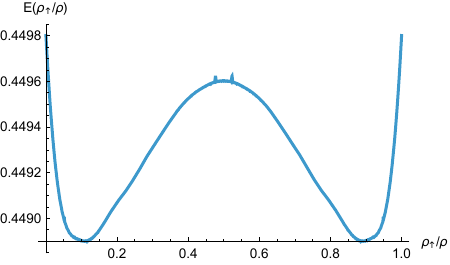}
        \caption{When we increase $\rho$, we observe that magnetization in the ground state approaches zero ($\rho=0.54$)}
    \end{subfigure}
    \\
    \begin{subfigure}{0.35\textwidth}
        \centering
        \includegraphics[width=\linewidth]{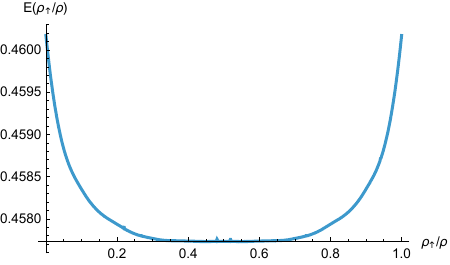}
        \caption{At some $\rho$ two minima connect to a single minimum with zero magnetization ($\rho=0.5445$)}
    \end{subfigure}
    \hspace*{1.5cm}
    \begin{subfigure}{0.35\textwidth}
        \centering
        \includegraphics[width=\linewidth]{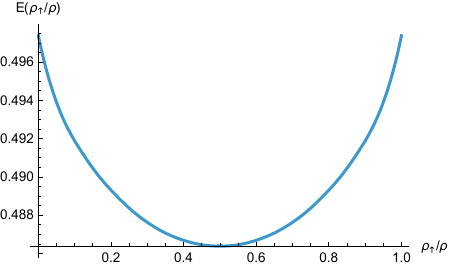}
        \caption{After increasing the $\rho$, minimum still has magnetization 0 ($\rho=0.56$)}
    \end{subfigure}
    
    \caption{Non-physical behavior for FCC lattice with $t_2 = t/2$ at $a = 0.16$ corresponding to $U = 5$}
    \label{fig:fcc05}
\end{figure}

We also present the results (presumably non-physical) for the case of maximal frustration for  $t_2=t/2$ (Fig. \ref{fig:fcc05}), which are analogous to the previous subsection, with the same explanation following from critical density of states and monotonicity of density of state in maximally frustrated case.

\begin{figure}[H]
    \centering
    \begin{subfigure}{0.35\textwidth}
        \centering
        \includegraphics[width=\linewidth]{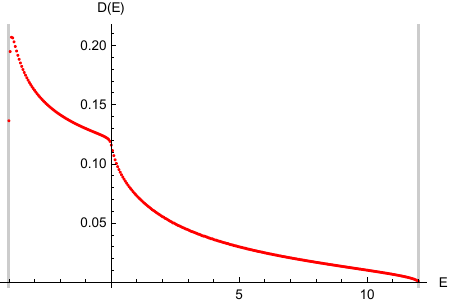}
    \end{subfigure}
    \hspace*{1.5cm}
    \begin{subfigure}{0.35\textwidth}
        \centering
        \includegraphics[width=\linewidth]{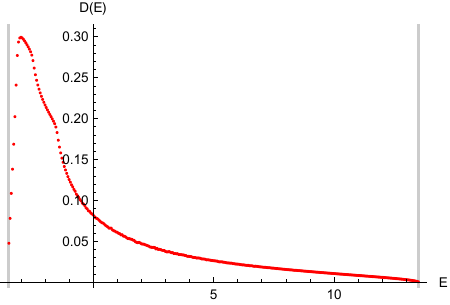}
    \end{subfigure}
    \\
    \begin{subfigure}{0.35\textwidth}
        \centering
        \includegraphics[width=\linewidth]{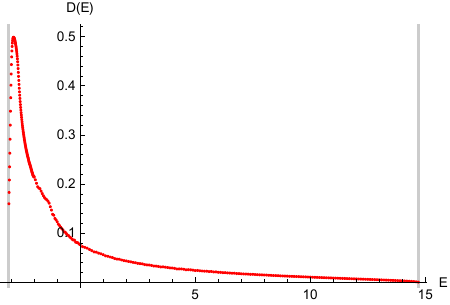}
    \end{subfigure}
    \hspace*{1.5cm}
    \begin{subfigure}{0.35\textwidth}
        \centering
        \includegraphics[width=\linewidth]{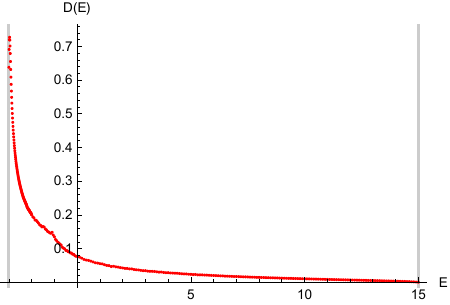}
    \end{subfigure}
    
    \caption{Density of states for the FCC lattice for $t=-1$ and $t_2=0, -0.25, -0.45, -0.5$ }
\end{figure}

Below we present summarized results concerning critical densities

\begin{table}[H]
	\centering
	\caption{Summarized results from Sec. \ref{sec:Main}}
	\begin{tabular}{|p{8cm}|c|c|c|}
		\hline\hline
		lattice & parameters ($t$ is always -1) & critical density $\rho_c$ & $\rho_c^{1/3} a$\\
		\hline\hline
		Simple cubic lattice & $U=20$ & no phase transition & $-$\\
		\hline
		Simple cubic lattice with NNN hoppings & $t_2=-0.25$, $U=20$ & $\rho_c=0.61$ & $0.26$\\
		\hline
		Simple cubic lattice with DNN hoppings & $t_2=-0.25$, $U=5$ & unphysical result & $-$\\
		\hline
		Simple cubic lattice with DNN hoppings, almost complete frustration & $t_2=-0.245$, $U=5$ & $\rho_c=0.031$ & $0.05$\\
		\hline
		FCC lattice, moderate frustration & $t_2=-0.25$, $U=5$ & $\rho_c=0.09$ & $0.07$\\
		\hline
		FCC lattice & $t_2=-0.5$, $U=5$ & unphysical result & $-$\\
		\hline\hline
	\end{tabular}
\end{table}

\section{Summary, perspectives of further research}
\label{sec:Summa}

Our paper is  focused on a study of the ground-state magnetization on
a variety of
 lattices with the use of low-density expansion.

To summarize the approach used: density expansion of ground-state energy possesses the structure:
\begin{align}
    E(\rho_+, \rho_-) = e_0(\rho_+) + e_0(\rho_-) + C\, a \,\rho_+\rho_- + {\cal{O}}(\rho_+, \rho_-)
\end{align}
($e_0$ -- energy as a function of density for non-interacting particles, $a$ -- scattering length).
The form of the ground-state energy is the same for both fermions in the continuum as well as the Hubbard model \cite{fermi exact}\cite{hub exact}.
The rigorous proof of an expression above is highly non-trivial, but the physics standing behind it is quite simple.
Namely, the properties of the system in the low-density region are determined by the two-particle scattering process.
It is natural to expect that the expression above, with only two terms, is a good approximation of energy for low densities.

Unfortunately, such expression didn't settle the presence of itinerant ferromagnetism in models analyzed so far: fermions in continuum and the Hubbard model on the simple cubic lattice.
For fermions in a continuum, the phase transition to a ferromagnetically ordered state happens at a density of order 1, i.e. far beyond the applicability of low-density expansion.
On the other hand, Giuliani proved that there is no ferromagnetism in the limit $\rho \to 0$; however, he didn't exclude the possibility of ferromagnetic ordering at some finite but low density.

In this paper,
 we analyzed the above expansion, neglecting the remainder term, for the Hubbard model on a few (five) lattices, including the simple cubic lattice and the (frustrated) FCC lattice.
We hoped to make more rigorous the qualitative statement, stated in the Stoner criterion, that 'frustration induces ferromagnetism'.
We expected that showing ferromagnetic ordering in those examples would be the starting point for exact results, showing that inclusion of higher-order terms will not change those qualitative results.

While with this method quite a large set of lattices (and interactions) can be analyzed, allowing for insight into 'wealthy' ferromagnetism, there are some limitations of our method.
It is limited to $d=3$ (for $d=1,2$ the low-density expansion has a completely different form and needs to be investigated from the beginning). 
Moreover, only low densities are allowed -- otherwise the low-density expansion loses its validity. 
But the fact that we have observed an occurrence of ferromagnetism in systems close to physical reality -- is an advantage of the method. 

In our findings, we have confirmed and made more quantitative the intuition that the frustration in the lattice aids ferromagnetic ordering.
The results show that while a simple cubic lattice shows no ferromagnetic ordering (unless we use very high unphysical values of the $U$ parameter), for frustrated lattices we observe a phase transition from a non-magnetic  ground state to a magnetic 
one.

The most important results concern FCC lattice (where phase transition takes place for moderate $U$ about 5 and low density about 0.1). 
This lattice describes nickel structure and therefore is of high physical relevance. 
Such a low value of critical density gives us hope that it will be possible to prove rigorously (for instance, by estimation that influence of higher-order terms will be negligible) an occurrence of itinerant 'wealthy' ferromagnetism, which -- to our best knowledge -- has not been proven so far. 
As an important step in this direction, it would be extremely interesting to compute higher-order terms of low-density expansion for the Hubbard model, independently of results obtained for fermions in continuum \cite{Kanno}, \cite{KannoRigorously}, \cite{ChWorder2}, \cite{ChWorder3}, and to analyze their impact on the presence of a phase transition and the emergence of ferromagnetic order.

There are many directions in which the setup can be generalized.
The most obvious one would be to study a more diverse set of lattices, in particular those that have multiband dispersion relations. 
Another interesting extension is to consider more complicated interaction between particles; instead of just interaction between particles on the same site, one can take into account interaction of particles on neighboring sites.
Consideration of non-zero temperature (beginning from the first order of density expansion and continuing to higher orders) would be the next natural circle of problems. 
Another includes examination of multiband Hubbard models and periodic Anderson models. 
Lastly, it would be interesting to compare our findings with experimental data. 
We hope that the low-density expansion can serve as a useful tool for interpreting results from experiments such as those reported in \cite{Young et al}, where ferromagnetism has been found for very low density.

{\bf Acknowledgments.} We are very indebted to Prof. Piotr Chankowski for numerous enlightening discussions.

\appendix
\appendixnumbering


\section{Exact formula for second derivative of total energy}
\label{sec:exact second derivative}
In the spirit of the Stoner criterion (see \cite{TasakiMB}), we can analyze formula \ref{low density} by computing the second derivative of the total energy with respect to the magnetization, defined as the ratio of spin-up particles to the total number of particles per site.
This analysis provides insight into the nature of the ground state: the sign of the second derivative indicates whether the system is in a local minimum or maximum with respect to magnetization.

Total energy as a function of the number of particles per site in the first order is

\begin{align}
    E_{tot}(\rho_\uparrow, \rho_\downarrow) = E_0(\rho_\uparrow) + E_0(\rho_\downarrow) + 8 \pi a \rho_\uparrow \rho_\downarrow \label{Etot(n)}
\end{align}

where $\rho_\sigma$ is the number of particles per site of spin $\sigma$, $E_0$ is the energy of the non-interacting system of identical particles, and $a$ is the scattering length. 
First, let us find the second derivative of $E_0$; we will use the fact that both $E_0$ and $\rho_\sigma$ can be calculated using the density of states $D(e)$ and integrating up to a certain Fermi energy $e_F$

\begin{subequations}
    \begin{align}
        \rho_\sigma(e_F) &= \int_{e_{min}}^{e_F} D(e) de\\
        E_0(e_F) &= \int_{e_{min}}^{e_F} e D(e) de
    \end{align}
\end{subequations}

As density of states is a positive function, $\rho_\sigma(e_F)$ is strictly increasing and can be inverted to $e_F(\rho_\sigma)$. The function $E_0(\rho_\sigma)$ is in fact the composition $E_0(e_F(\rho_\sigma))$, so using the chain rule and the formula for the derivative of an inverse function, we have 

\begin{align}
    \frac{d E_0}{d\rho_\sigma}(\rho_\sigma) &= \frac{d E_0}{d e_F}(e_F(\rho_\sigma)) \frac{d e_F}{d\rho_\sigma}(\rho_\sigma) = \frac{d E_0}{d e_F}(e_F(\rho_\sigma)) \left(\frac{d \rho_\sigma}{d e_F}(e_F(\rho_\sigma)) \right)^{-1} \\
    &= e_F(\rho_\sigma) D(e_F(\rho_\sigma)) (D(e_F(\rho_\sigma)))^{-1} = e_F(\rho_\sigma)
\end{align}

To find the second derivative, we apply the formula for the derivative of an inverse function again

\begin{align}
    \frac{d^2 E_0}{d\rho_\sigma^2}(\rho_\sigma) =  \frac{d e_F}{d\rho_\sigma}(\rho_\sigma) = \left(\frac{d \rho_\sigma}{d e_F}(e_F(\rho_\sigma)) \right)^{-1} = \frac{1}{D(e_F(\rho_\sigma))}
\end{align}

To analyze the ground state, we plotted $E_{tot}$ as a function of magnetization with fixed $\rho_{tot}=\rho_\uparrow+\rho_\downarrow$ and $a$, so we are interested in the second derivative of $E_{tot}$ over magnetization.
We introduce new variables

\begin{subequations}
    \begin{align}
        &\rho_{tot} = \rho_\uparrow+\rho_\downarrow &P = \frac{\rho_\uparrow}{\rho_{tot}}\\
        &\rho_\uparrow = \frac{1}{2}\rho_{tot}(1+P) &\rho_\downarrow = \frac{1}{2}\rho_{tot}(1-P)
    \end{align}
\end{subequations}

after rewriting $E_{tot}$ in new variables, we can compute derivatives

\begin{subequations}
    \begin{align}
        E_{tot} &= E_0\left(\frac{1}{2}\rho_{tot}(1+P)\right) + E_0\left(\frac{1}{2}\rho_{tot}(1-P)\right) + 2 \pi a \rho_{tot}^2 \left(1-P^2\right)\\
        \frac{d E_{tot}}{d P} &= \frac{1}{2}\rho_{tot} \left[e_F\left(\frac{1}{2}\rho_{tot}(1+P)\right) - e_F\left(\frac{1}{2}\rho_{tot}(1-P)\right)\right] - 4 \pi a \rho_{tot}^2 P\\
        \frac{d^2 E_{tot}}{d P^2} &= \frac{1}{4}\rho_{tot}^2 \left[\frac{1}{D\left(e_F\left(\frac{1}{2}\rho_{tot}(1+P)\right)\right)} +  \frac{1}{D\left(e_F\left(\frac{1}{2}\rho_{tot}(1-P)\right)\right)}\right] - 4 \pi a \rho_{tot}^2
    \end{align}
\end{subequations}

We are interested in the value of derivatives at P=0

\begin{subequations}
    \begin{align}
        \frac{d E_{tot}}{d P} \big|_{P=0}&= 0\\
        \frac{d^2 E_{tot}}{d P^2} \big|_{P=0}&= \frac{1}{2}\rho_{tot}^2 \frac{1}{D\left(e_F\left(\frac{1}{2}\rho_{tot}\right)\right)} - 4 \pi a \rho_{tot}^2 = \frac{1}{2}\rho_{tot}^2 \left[ \frac{1}{D\left(e_F\left(\frac{1}{2}\rho_{tot}\right)\right)} - 8 \pi a \right]
    \end{align}
\end{subequations}

The $\rho_{tot}^2$ term factors out in front of the expression, so the sign of the second derivative depends only on $\frac{1}{D}-8 \pi a$.
We can define critical density of states as 

\begin{align}
    D_{crit} =  \frac{1}{8 \pi a}
\end{align}

When the density of states corresponding to half of the total density of particles is smaller than the critical density of states, then at zero magnetization we will have a local minimum.
Unfortunately, we are unable to determine the magnetization of the ground state based on that, because the global minimum may be different, as seen in first-order phase transition in sections \ref{NNN} and \ref{DNN} (in non-physical case).
However, when the density of states is above the critical value, the ground state will always have non-zero magnetization, as at $P=0$ we will have a local maximum. 
Nevertheless, the above formula gives us insight into our results and phase transitions in the Hubbard model.
For simple a cubic lattice, we don't see a phase transition for $U=20$ and indeed the critical density is $0.176$, while the supremum of $D(e_F)$ is equal to $0.141$. 
Critical density can only be achieved if we increase $U$; to have critical density lower than $0.141$, $U$ needs to be bigger than 67 which is a non-physical value for this parameter.

Critical density of states also allows us to understand better the incorrect direction of phase transition in section \ref{DNN} and the FCC lattice at $t_2=t/2$.
In those cases density of states calculated from the Green's function is decreasing, so if function $D(e)$ and $y=D_{crit}$ intersect at some energy, then for energies below, density of states is bigger than critical value, so at magnetization zero we have a local maximum; therefore, the only type of phase transition that we can have with such density of state is magnetized $\rightarrow$ non-magnetized.
It must be noted that the FCC lattice should decrease to zero as energy goes to minimum as seen in \cite{FCC}; this behavior is not captured by numerical calculations of green functions that are not accurate near singularities.
If we could accurately compute the density of states, we would see this phase transition from a non-magnetized to a magnetized ground state.

This formula allows us finally to propose conditions that would allow for phase transition to occur in our system.
First we need densities large enough so that for some energy, density will be equal to $D_{crit}$ even for relatively small $U$, unlike in a simple cubic lattice, where we can only have a phase transition for very large and non-physical $U$.
A scenario that would guarantee this outcome is a singularity in the density of states.
Our formula is an approximation for small particle densities, so ideally we need this critical density to be reached for low energy so that the corresponding density of particles is as small as it can be.
This is why the FCC lattice in the $t_2=t/4$ case, where our density of states is better calculated, we see the correct phase transition for a relatively small density of particles and $U$.



\begin{thebibliography}{10}
 \bibitem{TasakiMB} H. Tasaki: Physics and Mathematics of Quantum Many-Body Systems. Springer 2020.
\bibitem{fermi exact} E. H. Lieb, R. Seiringer, J. P. Solovej, Ground State Energy of the Low Density Fermi Gas, 	Phys. Rev. A \textbf{71}, 053605 (2005).
\bibitem{hub exact} A. Giuliani, Ground state energy of the low density Hubbard model. An upper bound, J. Math. Phys. \textbf{48}, 023302 (2007).
    \bibitem{Mattis} D. C. Mattis, Theory of Magnetism Made Simple: An Introduction to Physical Concepts and to Some Useful Mathematical Methods, World Scientific Publishing Co. Pte. Ltd, Singapore 2006.
      \bibitem{Hubbard} J. Hubbard, Electron correlations in narrow energy bands, Proc. Roy. Soc. London A \textbf{276}, 238 (1963).
    \bibitem{Gutz} M.C. Gutzwiller, Effect of Correlation on the Ferromagnetism of Transition Metals, Phys. Rev. Lett. \textbf{10}, 159 (1963).
    \bibitem{Kana} J. Kanamori, Electron Correlation and Ferromagnetism of Transition Metals, Prog. Theor. Phys. \textbf{30}, 275 (1963).
    \bibitem{Stoner} E. Stoner, Lond. Edinb. Dublin Philos. Mag. J. Sci. \textbf{15}, 1018 (1933); Magnetism and Matter, Methuen, London 1934.
     \bibitem{Lenz} W. Lenz, Z. Phys. \textbf{56}, 778 (1929).
    \bibitem{Bloch} F. Bloch, Z. Phys. \textbf{61}, 206 (1930).
        \bibitem{Mielke}  A. Mielke, J. Phys. A: Math. Gen. \textbf{25}, 4335 (1992).
    \bibitem{Lieb ferri} E. H. Lieb, Phys. Rev. Lett. \textbf{62}, 1201 (1989).
    \bibitem{met-ins}M. Vojta, Rep. Prog. Phys. \textbf{66}, 2069 (2003).
    \bibitem{supercond} P.W. Anderson, The theory of superconductivity in the high-$T_c$ cuprates, Princeton Series in Physics (Princeton
    Univ. Press, 1997).
    \bibitem{cold} M. Aizenman , Phys. Rev. A \textbf{70}, 023612 (2004).
    \bibitem{nagaoka}  Y. Nagaoka, Phys. Rev. \textbf{147}, 392 (1966).
    \bibitem{tasaki} H. Tasaki, Ferromagnetism in the Hubbard model: A constructive approach, Comm. Math. Phys. \textbf{242}, 445-472 (2003).
\bibitem{TasakiAlmostFlat} H. Tasaki:  J. Stat. Phys. \textbf{84}, 535 (1996).
\bibitem{TanakaTasaki} A. Tanaka and H. Tasaki: Phys. Rev. Lett. \textbf{98}, 116402 (2006).
    \bibitem{tasaki more} H. Tasaki, The Hubbard Model: Introduction and Selected Rigorous Results, J. Phys.: Condens. Matter \textbf{10} 4353 (1998).
       \bibitem{Lieb} E. Lieb, The Hubbard Model: Some Rigorous Results and Open Problems, XI Int. Cong. MP, Int. Press 392-412 (1995).
    \bibitem{LeeYang} T. D. Lee and C. N. Yang: Phys. Rev. \textbf{105}, 1119 (1957).
        \bibitem{BachGrafSolovej} V. Bach, E. H. Lieb and J. P. Solovej,  Generalized Hartree-Fock theory and the Hubbard model, J. Stat. Phys. \textbf{76}, 3-89 (1994).
    \bibitem{Hirsch} J. E. Hirsch and S. Tang, Phys. Rev. Lett. {\bf 62}, 591 (1989).
    \bibitem{fermi approx1} K. Huang, C.N. Yang, Quantum-Mechanical Many-Body Problem with Hard-Sphere Interaction, Phys. Rev. \textbf{105}, 767-775 (1957)
    \bibitem{fermi approx2} T.D. Lee, C.N. Yang, Many-Body Problem in Quantum Mechanics and Quantum Statistical Mechanics, Phys. Rev. \textbf{105}, 1119-1120 (1957)
    \bibitem{fermi approx3} A.L. Fetter, J.D. Walecka, Quantum Theory of Many-Particle Systems, McGraw-Hill, New York (1971)
    \bibitem{FermiExact2} M. Falconi, E. L. Giacomelli, C. Hainzl, and M. Porta, The dilute Fermi gas via Bogoliubov theory. Ann. Henri Poincar\'{e} 22, 2283-2353 (2021)
    \bibitem{Kanno} S. Kanno, Prog. Theor. Phys. \textbf{44}, 813 (1970)
    \bibitem{ChWorder2} P. H. Chankowski, J. Wojtkiewicz, Phys. Rev. \textbf{B} 104,
    144425 (2021)
    \bibitem{ChWorder3} P. H. Chankowski, J. Wojtkiewicz and S. Augustynowicz, Phys. Rev. \textbf{A} 107, 063311 (2023).
    \bibitem{KannoRigorously} E. L. Giacomelli, Ch. Hainzl, P. T. Nam and R. Seiringer: Comm. Pure Appl. Math. 2026, 0:1-54.
        \bibitem{SeiringerYin} R. Seiringer and J. Yin, J. Stat. Phys. {\bf 133}, 1139 (2008)
    \bibitem{DerGer} J. Derezi\'{n}ski and C. G\'{e}rard: Scattering Theory of Classical and Quantum N-Particle Systems. Springer, 1997.
    \bibitem{Obermeier_DMFT} T. Obermeier, T. Pruschke and J. Koller: Ferromagnetism in the large-$U$ Hubbard model. Phys. Rev. B 56, R8479(R) (1997)
    \bibitem{KotliarRuckenstein} G. Kotliar and A. E. Ruckenstein, Phys. Rev. Lett. {\bf 57}, 1362 (1986)
    \bibitem{economu} E. N. Economou,  Green's Functions in Quantum Physics, Springer (2006)
    \bibitem{hub cubic} T. Hanisch, G. S. Uhrig, and E. Müller-Hartmann, Lattice dependence of saturated ferromagnetism in the Hubbard model, Phys. Rev. B \textbf{56}, 13960 (1997)
    \bibitem{FCC} D. Vollhardt, N. Blümer, K. Held, M. Kollar, Metallic Ferromagnetism -- an Electronic Correlation Phenomenon.
Lecture Notes in Physics, Vol. 580, (Springer, Heidelberg, 2001) p. 191-207.
    \bibitem{Lieb93} Lieb, E. H. In: Proceedings for Advances in Dynamical Systems and Quantum Physics, Capri, 1993 World Scientfic, River Edge, NJ, 1995, pp. 173-193.
    \bibitem{Young et al} D. Young, D. Hall, M. Torelli et al., High-temperature weak ferromagnetism in a low-density free-electron gas, Nature \textbf{397}, 412-414 (1999).
\end{thebibliography}
\end{document}